\documentclass[11pt]{article}
\usepackage{epsfig,amssymb,graphicx,amsmath,amsthm,subcaption}
\usepackage{authblk}
\usepackage{array}

\newtheorem{remark}{Remark}
\newtheorem{lemma}{Lemma}

\newtheorem{theorem}{Theorem}

\begin{document}

\title{Modelling the effects of awareness-based interventions to control the mosaic disease of {\it Jatropha curcas}}
\author{Fahad Al Basir$^1$, Konstantin B. Blyuss$^{\rm 2}$\footnote{Corresponding author: K.Blyuss@sussex.ac.uk}, Santanu Ray$^1$
}

\affil{$^{\rm 1}$ Systems Ecology \& Ecological Modeling Laboratory, Department of Zoology, Visva-Bharati University, West Bengal - 731235, India}

\affil{$^{\rm 2}$ Department of Mathematics, University of Sussex, Falmer, Brighton BN1 9QH, UK}

\maketitle

\begin{abstract}

Plant diseases are responsible for substantial and sometimes devastating economic and societal costs and thus are a major limiting factor for stable and sustainable agricultural production. Diseases of crops are particular crippling in developing countries that are heavily dependent on agriculture for food security and income. Various techniques have been developed to reduce the negative impact of plant diseases and eliminate the associated parasites, but the success of these approaches strongly depends on population awareness and the degree of engagement with disease control and prevention programs. In this paper we derive and analyse a mathematical model of mosaic disease of {\it Jatropha curcas}, an important biofuel plant, with particular emphasis on the effects of interventions in the form of nutrients and insecticides, whose use depends on the level of population awareness. Two contributions to disease awareness are considered in the model: global awareness campaigns, and awareness from observing infected plants. All steady states of the model are found, and their stability is analysed in terms of system parameters. We identify parameter regions associated with eradication of disease, stable endemic infection, and periodic oscillations in the level of infection. Analytical results are supported by numerical simulations that illustrate the behaviour of the model in different dynamical regimes. Implications of theoretical results for practical implementation of disease control are discussed.


\end{abstract}

\section{Introduction}

Constantly increasing global energy demands have significantly raised the need for stable alternative fuel sources. One the most prominent types of alternative energy is the biofuels that are produced from oils of a variety of plants, many of which can be grown in a sustainable manner even in harsh environmental conditions. Among various candidates for the mass production of biofuel, \textit{Jatropha curcas} has recently emerged as a strong contender, due to its high content of 27-40\% of triglycerides \cite{ach,Sah}, and the fact that this plant can be grown even in drought conditions, on arid, salty and sandy soils, it requires minimum cultivation efforts and produces first harvest in just 18 months. Moreover, the reported levels of oil production from Jatropha plants are higher than those of soybean (the main source of biodiesel in the US), sesame, sunflower, castor and rapeseed from plantations of the same size \cite{Jong}. The Jatropha plant originated in Central America and Mexico, but has subsequently spread to Africa, Latin America and South-East Asia. Importantly, the Jatropha plant does not compete with other food crops, and beside being a source of biofuel, it also proves to be an effective phytoremediator, carbon sequester, and soil erosion controller \cite{mang08,pand12}.

A major challenge for the sustainable large-scale growth of the Jatropha is plant disease \cite{security,Africa,virus diseases}, most often a mosaic disease caused by one of the viruses in the {\it Begomovirus} family \cite{Kash, Nar,Gao,Reddy} that is transmitted by the whitefly {\it Bemisia tabaci} \cite{Bedford}. The effects of this disease include mosaiced, reduced and distorted leaves, blistering, as well as stunting of diseased plants. Low density of \textit{Jatropha curcas} is known to facilitate fast transmission of mosaic disease \cite{Fauq}, and the disease transmission is affected by environmental conditions such as temperature and humidity, with heavy rainfalls significantly limiting the spread of whiteflies \cite{Fargette}. The virus is transmitted from infected plants to uninfected vectors, and from infected vectors to uninfected plants. Once the vectors acquire mosaic virus from infected plants, they are able to pass it on to other uninfected plants within 48 hours \cite{Fargette}.

Various strategies have been developed to mitigate the negative effects of mosaic disease \cite{strategy,strategy2,Ahohuendo}. These include vector control in the form of insecticidal soaps \cite{soap,Thresh}, as well as application of nutrients to the soil. Insecticidal soaps are sprayable organic insecticides that can be used on a variety of plants, fruit and vegetables, to a degree that these products can be safely consumed after normal washing. Their insecticidal action consists in blocking the spread of whitefly-borne infection by reducing the number of eggs being laid, as well as preventing adults from flying, thus minimising the disk of further disease transmission. Insecticidal soaps have already proved to be effective in reducing pest infections of cottonseed and cowpea \cite{Butler,Opa}. Another effective approach for control of mosaic disease is the use of nutrients that can reduce disease burden by providing disease tolerance or resistance of plants to pathogens \cite{nutrient,nutrient1,1970,Graham1,Graham2}. Plant nutrition is an essential component of sustainable agriculture, as in most cases it is more cost-effective and also environmentally friendly to control plant disease with the adequate amount of nutrients without the use of pesticides. Once the level of disease is reduced to an appropriate level, it can be further controlled by other cultural practices or conventional organic biocides, making this approach not only successful, but also less expensive. There are several examples of efficient disease control through manipulation of soil nutrient concentration, which can be achieved by modifying either nutrient availability, or nutrient uptake \cite{Huber1999}.

Most effective strategies for control of plant disease include a combination of different approaches, as in integrated pest management \cite{Schut,Klerkx,aware_knowledge1}. It should be noted, however, that a successful implementation of a large-scale crop disease containment and prevention program can only be achieved subject to adequate level of population awareness and cooperation \cite{aware_plant1,Khan}. This would not only improve the uptake of cultivating a particular crop by farmers, but also would facilitate their engagement in improving crop performance and disease control \cite{LeB}.  Farming awareness campaign in Malenadu region in India helped educate farmers on the serious risks that pesticides pose both to the human health and to the environment, and to encourage proper use of pesticide to minimise their negative effects \cite{Kumar2,Yang2014}. Similar approach was used in Indonesia, where dedicated farmer field schools were used to disseminate information about sensible farming practices that resulted in improved cost-effectiveness and reduced unnecessary use of pesticides \cite{school,school2,field_schools}. In the particular case of cultivating Jatropha plants for the purpose of developing additional income from biofuel, major information campaigns in Kenya by various NGOs, community-based organisations and private investors, have led to the large-scale adoption of {\it J. curcas} by farmers \cite{Mogaka}. Mali has designed a dedicated governmental Strategy for Biofuels Development aimed at promoting {\it J. curcas} as a sustainable development
tool \cite{Favr}. In Burma, the national campaign for biodiesel production took off on an unprecedented scale in 2005, with funds, farm lands and labour being diverted to growing Jatropha \cite{WRM08}. From the perspective of responding to mosaic disease, proactive involvement of farmers has proved very effective in improving disease control and subsequently increasing crop yields \cite{aware_mosaic1,aware_mosaic2}.

A number of mathematical models have looked at effects of population awareness on control of infectious diseases \cite{Misra1,Misra2,ManOn,Cui,Funk,agaba1}. Time delay associated with response to disease awareness has also been shown to play a significant role in determining disease outcome and design of appropriate control measures \cite{greenh,zuo,agaba2,agaba3}. In terms of modelling the effects of awareness on control of mosaic disease in {\it J. curcas}, Al Basir et al. \cite{mmas} analysed a model with a separate compartment for aware population, and assumed that removal of infected plants and infected vectors occurs at a rate proportional to the number of aware individuals. Al Basir and Roy \cite{ABR} studied the effect of roguing, i.e. removing of infected plants, at a rate proportional to the overall number of infected plants, with a time delay to account for the time it takes to observe the infection and take action. Without making it explicit, effectively this represents the response of farmers through their delayed awareness of mosaic disease affecting Jatropha plants. Roy et al. \cite{cmbn} have analysed a model of mosaic diseased and used significant similarities between mosaic infections of cassava and Jatropha plants to parameterise their model and investigate the impact of continuous and pulse spraying strategies for the application of insecticidal soap to eliminate vector population. Venturino et al. \cite{eee} have considered the same problem with continuous spraying from the perspective of optimisation theory and showed how an optimal strategy can be developed that minimises the use of insecticide, while achieving the aim of controlling the spread of mosaic disease.

In this paper we consider the spread of mosaic disease in a Jatropha plantation, with disease control being implemented through the application of insecticides and nutrients depending on the level of population awareness about the disease. The awareness is assumed to have a contribution from direct observation of plant infection by farmers, and another input from global awareness campaigns. The outline of this paper is as follows. In the next section we derive the mathematical model of mosaic disease of Jatropha plants and discuss its basic properties. Section 3 is devoted to analysis of stability and bifurcation of different steady states of the model. In Section 4, we supplement analytical results by numerical computation of bifurcation diagrams, as well as numerical solution of the model to illustrate different dynamical regimes. The paper concludes in Section 5 with the discussion of results and future research.

\section{Model derivation}

We consider a population of plants that can become exposed to a mosaic disease spread by a whitefly vector. Plant population is divided into healthy, latently infected, and infected plants, to be denoted as $x$, $l$ and $y$, respectively. Healthy plants are assumed to reproduce logistically with a growth rate $r$ and a carrying capacity $K$. It is assumed that once whiteflies infect a healthy plant, it becomes latently infected, i.e. it is incubating the disease but does further contribute to new infections. Rather than explicitly model the process of transfer of infection from plants to vectors, we instead focus directly on the population of infected vectors, whose size is denoted by $v$, and assume that the rate of growth of infected vectors is proportional with a constant $b$ to the number of infected plants, from which they can acquire the infection. Begomoviruses that cause mosaic disease are known to be {\it circulative-persistent viruses} \cite{Vir}, which means that once the whitefly vectors become infected, they will remain infectious for the rest of their lifetime \cite{Holt97,Jeger04}. The reason for this is that when whiteflies feed on infected plants, they ingest the virus contained in the plant sap with their stylets, and subsequently the virus crosses the filter chamber and the midgut to be then translocated into the primary salivary glands \cite{Vir,Jackson}. When these vectors then feed on healthy plants, virus particles circulating in the whitefly saliva will enter these plants and start infection in them.

With these assumptions, the basic host-vector model for the dynamics of mosaic disease takes the form
\begin{equation}\label{mod1}
\begin{array}{l}
\displaystyle{\frac{dx}{dt}=r x\left( 1-\frac{x+l+y}{K}\right) - \lambda x v,}\\\\
\displaystyle{\frac{dl}{dt}=\lambda x v-\alpha l,}\\\\
\displaystyle{\frac{dy}{dt}=\alpha l - m y,}\\\\
\displaystyle{\frac{dv}{dt}=by - \eta v,}
\end{array}
\end{equation}
where $1/\alpha$ is the average period of latency, after which plants start to exhibit symptoms of infection and are capable of passing the infection to uninfected vectors, $m$ is the death rate of infected plants, $1/\eta$ is the average lifetime of infected vectors, and $\lambda$ is the rate of infection of healthy plants by infected vectors.

All interventions targeting the spread of mosaic disease stem from being aware that this disease is actually affecting plant population. If we denote by $M(t)$ the level of population awareness of the mosaic disease, its dynamics is then described by the equation
\begin{equation}
\frac{dM}{dt}=\omega_0+\sigma y-\tau M,
\end{equation}
where $\omega_0$ is the rate of global awareness due to media campaigns etc., $\sigma$ is the increase of awareness due to observation of infected plants, and $\tau$ is the rate at which awareness dissipates ($1/\tau$ is the average duration of ``remembering'' about the disease). Increase in population awareness can result in two types of interventions: application of nutrients and/or the use of insecticide.

The amount of nutrients being used can be taken to increase with the level of awareness according to \cite{greenh,Misra11}
\begin{eqnarray}
N(M)=\mu+\frac{\rho M}{1+M},
\end{eqnarray}
where $\mu$ represents the level of nutrient use in the absence of any information about the disease, which can be related to the cost of nutrients being used, and $\rho$ denotes the maximum rate at which nutrient is applied in the plantation. This functional dependence is chosen in a way where if there is no awareness about the plant disease, depending on their financial abilities, the farmers would be applying some small background amount of nutrients $\mu$ to improve plant performance. Once they become aware of the ongoing plant infection, i.e. for small values of population awareness $M$, they will increase the use of nutrients to protect their plants, and thus the level of application of nutrients will grow with $M$. However, as the awareness increases, eventually the amount of nutrients being applied saturates, as by that time all farmers are already fully aware of the plant infection and are using maximum available amount of nutrients to protect their plants.

The effect of using nutrients is two-fold: it facilitates a better/faster growth of healthy plants, and impedes the spread of infection by making healthy plants more resistant. The first of these effects can be incorporated in the above model by modifying the linear growth rate to become
\begin{eqnarray}
r(N)=r_0[1+k_1N(1-k_2N)],
\end{eqnarray}
so that the nutrients are beneficial when used in small quantities, but they can lead to plant
deficiency and cause plant death due to toxicity when large amounts of nutrients are applied \cite{nutrient,nutri}. The second effect of using nutrients can be formalised as follows,
\begin{eqnarray}
\lambda(N)=\frac{\lambda_0}{1-k_3e^{-k_4N}},~0\leq k_3<1,
\end{eqnarray}
which means that applying nutrients reduces the capacity of mosaic disease to infect healthy plants.

Another type of intervention that farmers can make when they become aware of the ongoing
mosaic disease is to use insecticides. These act to kill infected vectors, so the last equation of model (\ref{mod1}) becomes modified
\begin{equation}
\frac{dv}{dt}= by - \eta v-P(M)v ,
\end{equation}
where the function $P(M)$ quantifies how the rate of use of insecticide depends on the level of
awareness. Assuming there is a limit on how much or how quickly the insecticide can be used (due to logistical or financial constraints), this
function can be taken in the form
\begin{equation}
P(M)=\epsilon\frac{M}{1+M},
\end{equation}
where $\epsilon$ denotes the maximum level of insecticide use. With $M$ being constant, one would have $P(M)= \epsilon M/(1+M)=\gamma$, resulting in the term $-\gamma v$ in the equation for infected vectors, which is identical to a constant pesticide spraying strategy studied in earlier papers \cite{cmbn,eee}.

With these additional assumptions on possible interventions, the complete model has the form
\begin{equation}\label{model2}
\begin{array}{l}
\displaystyle{\frac{dx}{dt}= r(M) x\left( 1-\frac{x+l+y}{K}\right) - \lambda(M) x v,}\\\\
\displaystyle{\frac{dl}{dt}=\lambda(M) x v-\alpha l,}\\\\
\displaystyle{\frac{dy}{dt}=\alpha l - m y,}\\\\
\displaystyle{\frac{dv}{dt}=by - \eta v-P(M)v,}\\\\
\displaystyle{\frac{dM}{dt}=\omega_0+\sigma y-\tau M,}
\end{array}
\end{equation}
with $r(M)=r[N(M)]$ and $\lambda(M)=\lambda[N(M)]$, and the initial conditions: $x(0)>0,l(0)\geq 0,y(0)=0,v(0)>0,M(0)\geq 0$. The following region of the phase space
\[
\mathcal{D}=\left\{(x, l, y, v, M) \in \mathbb{R}^5_+ : 0\leq x,l,y\leq K, 0\leq v\leq \frac{bK}{\eta},0\leq M\leq \frac{\omega_0+\sigma K}{\tau}\right\} \]
is positively invariant, and it attracts all solutions with non-negative initial conditions.

\section{Equilibria and their stability}

For any parameter values, the system (\ref{model2}) has an awareness-only equilibrium $\displaystyle{E_0(0,0,0,0,\omega_0/\tau)}$ and a disease-free equilibrium $\displaystyle{E_1(K,0,0,0,\omega_0/\tau)}$. It can also have an endemic equilibrium $E^*(x^*,l^*,y^*,v^*,M^*)$ with
\begin{eqnarray*}
  x^* &=& \frac{m\left[\eta+P(M^*)\right]}{\lambda b},\quad
  l^* =\frac{m(\tau M^*-\omega_0)}{\alpha\sigma},\\\\
  y^* &=& \frac{\tau M^*-\omega_0}{\sigma},\quad
  v^*=\frac{b(\tau M^*-\omega_0)}{\sigma[\eta+P(M^*)]},
\end{eqnarray*}
and $M^*$ being a positive root of the following equation
\begin{eqnarray}
F(M)&=&r(M)\Big\{\sigma\left[\eta+P(M)\right]\cdot\left[K-\frac{\eta+P(M)}{\lambda(M)b}-\frac{(\tau M-\omega_0)(\alpha+m)}{\alpha\sigma}\right]\nonumber\\
&&+\lambda(M) Kb(\tau M-\omega_0)\Big\}=0.
\end{eqnarray}

Stability of each steady state $\bar{E}(\bar{x},\bar{l},\bar{y},\bar{v},\bar{M})$ is determined by the eigenvalues of the Jacobian evaluated at that steady state,
\[
J_{\bar{E}}= \left[\begin{array}{ccccc}
 J_{11}\quad & \displaystyle{-\frac{r(\bar{M})\bar{x}}{K}}\quad& \displaystyle{-\frac{r(\bar{M})\bar{x}}{K}}& -\lambda(\bar{M})\bar{x} &\quad J_{15}\\
\lambda(\bar{M})\bar{v}\quad & -\alpha \quad& 0 &\lambda(\bar{M})\bar{x} &\quad \lambda'(\bar{M})\bar{x}\bar{v}\\
0 \quad& \alpha \quad& -m & 0 &\quad 0\\
0 \quad& 0 \quad& b & -\eta-P(\bar{M}) & \quad-P'(\bar{M})\bar{v}\\
0 \quad& 0 \quad& \sigma & 0& \quad-\tau\\
\end{array}\right],
\]
where,
\[
\begin{array}{l}
\displaystyle{J_{11}=r(\bar{M})\Big(1-\frac{2\bar{x}+\bar{l}+\bar{y}}{K}\Big)-\lambda(\bar{M})\bar{v},\quad \!\!\!
\lambda'(\bar{M})=\lambda(\bar{M})\left[\frac{\rho k_3k_4e^{-k_4N(\bar{M})}}{(1+\bar{M})^2(k_3e^{-k_4\bar{M}}-1)}\right]\!,}\\\\
\displaystyle{J_{15}=r'(\bar{M})\bar{x}\left[1-\frac{(\bar{x}+\bar{l}+\bar{y})}{K}\right]-\lambda'(\bar{M})\bar{x}\bar{v},\quad P'(\bar{M})=\frac{\epsilon}{(1+\bar{M})^2},}\\\\
\displaystyle{r'(\bar{M})=r_0k_1\left[\frac{\rho}{(1+\bar{M})^2}[1-2k_2N(\bar{M})]\right].}
\end{array}
\]
Awareness-only equilibrium, $E_0$, always has a real positive eigenvalue, and hence, is always unstable. At the disease-free equilibrium $E_1$, the characteristic equation for eigenvalues $\xi$ factorises as follows,
\[
\begin{array}{l}
(\tau+\xi)\cdot[r(M_0) + \xi]\cdot\Big[\xi^3 + \xi \alpha (P(M_0) + \eta]) +
   m (\xi + \alpha) (\xi + P(M_0) + \eta])\\
   \hspace{1cm}+\xi^2 (\alpha + P(M_0) + \eta) -   b \alpha K \lambda(M_0)\Big]=0,\quad M_0=\omega_0/\tau.
   \end{array}
 \]
If we define the basic reproduction number $R_0$ as
\begin{equation}\label{R0def}
\displaystyle{R_0=\frac{bK\lambda(\omega_0/\tau)}{m\left[\eta+P(\omega_0/\tau)\right]}=\frac{bK(\tau+\omega_0)\lambda_0}{m[\eta\tau+\omega_0(\epsilon+\eta)]}\cdot\frac{1}{1-k_3\exp\left[-\frac{k_4[\mu\tau+(\mu+\rho)\omega_0]}{\omega_0+\tau}\right]},}
\end{equation}
then we have the following result.
\begin{theorem}\label{Thm1}
Disease-free equilibrium $E_1$ is stable if $R_0<1$, unstable if $R_0>1$, and undergoes a transcritical bifurcation at $R_0=1$.
\end{theorem}

\begin{remark}\label{Rem1}
It is important to note that $R_0$ does not depend on $\sigma$, hence, irrespective of how efficiently farmers become aware of disease due to observation of infected plants, this by itself is not sufficient to result in the eradication of infection. At the same time, taking a limit of $\omega_0\to\infty$ gives
\[
\lim_{\omega_0\to\infty}R_0(\omega_0)=R_0^{\infty}\equiv\frac{bK\lambda_0}{m(\eta+\epsilon)}\cdot\frac{1}{1-k_3\exp[-k_4(\mu+\rho)]}.
\]
In light of the fact that $R_0$ is monotonically decreasing with increasing $\omega_0$, this suggests that eradication of mosaic disease, as represented by a stable disease-free steady state $E_1$ is only possible if $R_0^{\infty}<1$, and since $0\leq k_3<1$, the only available means to achieve this is by increasing the rate of use of insecticide $\epsilon$.
\end{remark}

Characteristic equation at the endemic equilibrium $E^*$ has the form
 \begin{eqnarray}\label{char}
 \xi^5+A_1 \xi^4 + A_2 \xi^3 + A_3 \xi^2 + A_4 X +A_5=0,
\end{eqnarray}
with
\[
\begin{array}{l}
A_1=\frac{rx^*}{K} + \eta+P(M^*) + m + \alpha + \tau,\\\\
\displaystyle{A_2=\frac{r x^*}{K}[\lambda(M^*)v^*+m+\alpha+\tau+\eta+P(M^*)]+[\eta+P(M^*)](m+\alpha-\tau)}\\
\hspace{1cm}+m(\alpha+\tau)+\alpha\tau,\\\\
\displaystyle{A_3=\frac{r x^*}{K}\Big[(m+\alpha+\tau)[\eta+P(M^*)+\lambda(M^*) v^*]+\lambda(M^*) v^*[\eta+P(M^*)]\Big]}\\
\displaystyle{+\left(\eta+P(M^*)+\frac{r x^*}{K}\right)[m(\alpha+\tau)+\alpha\tau]+\alpha\Big[m\tau-x^*(b(M^*)+\sigma\lambda'(M^*)v^*)\Big],}\\\\
\displaystyle{A_4=\frac{r x^*}{K}\Bigg\{[\eta+P(M^*)]\left[\lambda(M^*) v^*(m+\alpha+\tau)+m(\alpha+\tau)+\alpha\tau\right]}\\
\displaystyle{+\tau\lambda(M^*) v^*(m+\alpha)+\alpha(m\tau-b\lambda(M^*)x^*)+\alpha\sigma\lambda'(M^*)x^*v^*\Bigg\}}\\
+\alpha[\eta+P(M^*)](m\tau-\sigma\lambda'(M^*)x^*v^*)+\alpha\lambda(M^*)x^*[v^*(b\lambda(M^*)+\sigma P'(M^*))-b\tau],
\\\\
\displaystyle{A_5=\frac{r x^*\tau}{K}[\eta+P(M^*)]\cdot[(m+\alpha)\lambda(M^*)v^*+m\alpha]+\frac{r \alpha (x^*)^2}{K}\Big[\sigma\lambda(M^*)v^*P'(M^*)}\\
\displaystyle{-b\tau\lambda(M^*)-[\eta+P(M^*)]\lambda'(M^*)v^*\Big]+\alpha x^*v^*\lambda^{2}(M^*)[b\tau-\sigma v^*P'(M^*)]}\\
+\alpha \sigma J^*_{15} v^*\lambda(M^*)  [\eta+P(M^*)].
\end{array}
\]
According to the Routh-Hurwitz criterion, all roots of the characteristic equation \eqref{char} have a negative real parts if the following conditions hold
\begin{equation}\label{RHstab}
\begin{array}{l}
A_1>0,\quad A_5>0, \quad A_1A_2-A_3>0, \quad A_3(A_1A_2-A_3)-A_1(A_1A_4-A_5)>0,\\\\
(A_1A_2-A_3)(A_3A_4-A_2A_5)+(A_1A_4-A_5)(A_5-A_1A_4)>0.
\end{array}
\end{equation}

Since the endemic steady state $E^*$ of the model depends on a large number of parameters, we use the general methodology developed in \cite{Dous15} for a complete characterisation of a Hopf bifurcation in a five-dimensional phase space. In principle, any of the parameters can be considered as a bifurcation parameter while other parameters are fixed, so we have the following result for any such parameter $\zeta$.

\begin{figure}
  \hspace{-1cm}
  \includegraphics[width=7in]{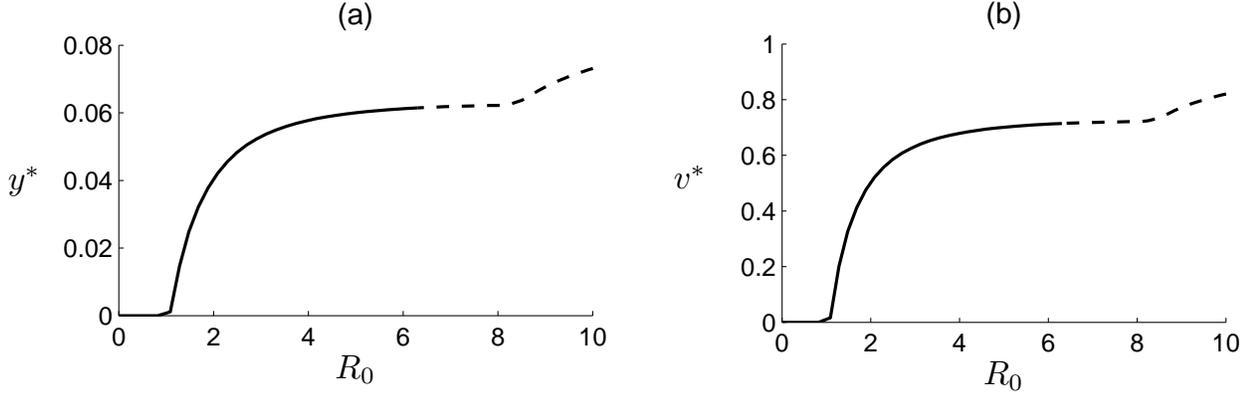}
  \vspace{-0.5cm}
  \caption{Steady state values of the infected plant biomass $y^*$ and infected vector $v^*$ population depending on the basic reproduction
  number $R_0$. The parameter values are as follows, $r_0=0.05,k_1=k_2=k_3=k_4=0.5, \mu=0.5, \rho=0.5, K=1, \alpha=0.1, b=0.8, \zeta=0.05,\epsilon=0.25, m=0.1, s=0.6,\sigma=0.05, \eta=0.05, \tau=0.05, \omega_0=0.001.$}\label{fig1}
\vspace{-0.3cm}
\end{figure}

\begin{theorem}\label{Hopf_thm}
The endemic equilibrium $E^*$ is stable if the conditions (\ref{RHstab}) hold. At $\zeta=\zeta^*$, the steady state $E^*$ undergoes a Hopf bifurcation, if either of the conditions below is satisfied.
\begin{itemize}
\item[i.]$\Psi(\zeta^*)=0$ {\rm and} $\displaystyle{\frac{d\Psi}{d\zeta}\Big|_{\zeta=\zeta^*}\neq 0}$, {\rm where} $$
\Psi(\zeta)=(A_3-A_1A_2)(A_5A_2-A_3A_4)-(A_5-A_1A_4)^2,
$$
with
\[
\theta=\frac{A_5-A_1A_4}{A_3-A_1A_2}>0,\quad A_3-A_1\theta\neq 0,
\]

\item[ii.] $A_5=A_1A_4$, $A_3=A_1A_2$, $A_4<0$, $A_1A_3\neq 0$,
\[
\Big[A_1'\theta^2+(A_1A_2'-A_3')\theta-(A_1A_4'-A_5')\Big]\Big|_{\zeta=\zeta^*}\neq 0.
\]
with
\[
\theta=\frac{1}{2}\left(A_2+\sqrt{A_2^2-4A_4}\right)>0.
\]
\end{itemize}
\end{theorem}

\noindent The proof of this theorem is given in \textbf{Appendix A}.

\section{Numerical stability analysis and simulations}\label{num}

To gain a better understanding of how different parameters affect the dynamics, in this section we investigate the stability of the steady states numerically, and also solve the system (\ref{model2}) to illustrate different types of behaviour. Parameter values are mostly taken from \cite{mmas,Holt97}, with the rest of them being hypothetical/estimated. Figure~\ref{fig1} shows how the steady state values of the infected plant biomass and infected vector population at the endemic steady state vary with the basic reproduction number. When $R_0<1$, only the disease-free steady state is feasible and stable, and at $R_0=1$, the disease-free steady states loses its stability via a transcritical bifurcation, and the stable endemic steady state $E^*$ appears. As $R_0$ increases further, the endemic state can also lose its stability via Hopf bifurcation,
\newpage
\begin{figure}[h]
\hspace{-0.3cm}
  \centering
  \includegraphics[width=7in]{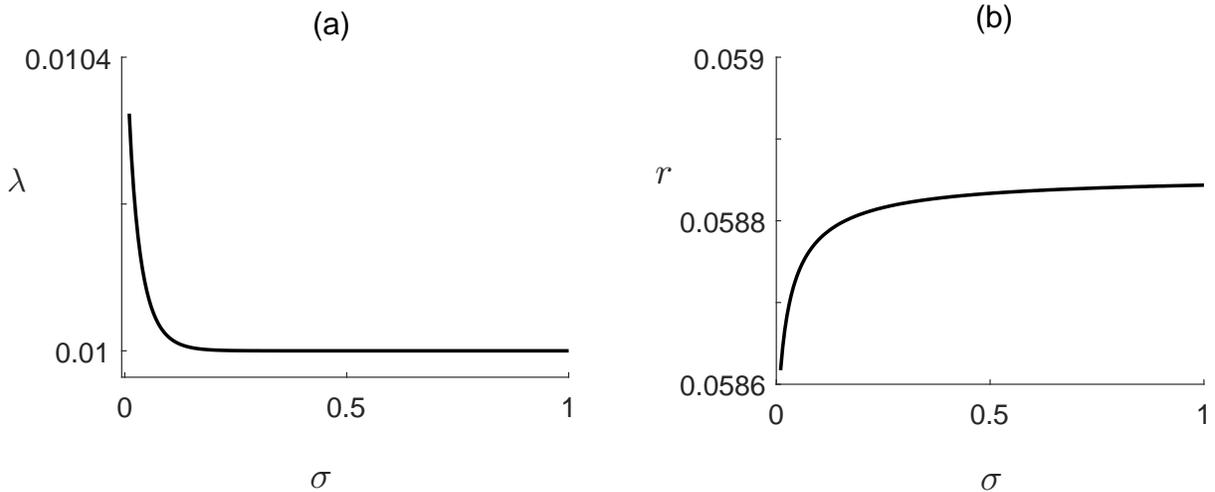}
  \vspace{-0.5cm}
  \caption{Dependence of the transmission rate $\lambda$ and growth rate $r$ on the rate of awareness $\sigma$ at the endemic steady state $E^*$. Parameter values are the same as in Fig.~\ref{fig1}.}\label{fig2}
  \vspace{-0.3cm}
\end{figure}
\begin{figure}[h]
\hspace{-1cm}
  \centering
  \includegraphics[width=7in]{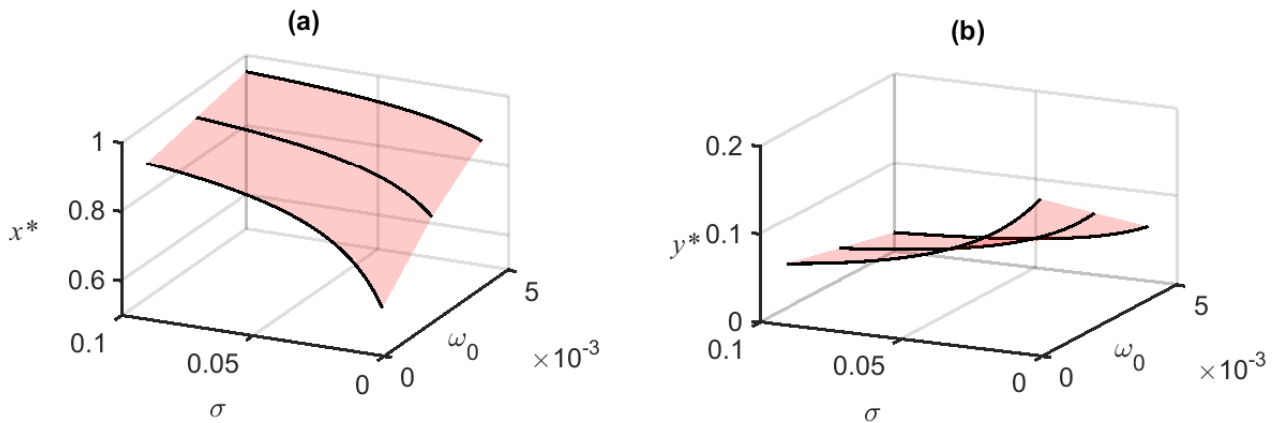}
  \vspace{-0.8cm}
  \caption{Dependence of the steady state values of the healthy and infected plant populations on the levels of global awareness $\omega_0$ and the awareness from observing infected plants $\sigma$. Parameter values are the same as in Fig.~\ref{fig1} except for $\lambda_0=0.003$.}\label{fig3}
  \vspace{-0.2cm}
\end{figure}

\noindent in agreement with {\bf Theorem~\ref{Hopf_thm}}. With an expression for $R_0$ given in (\ref{R0def}), to plot this figure we have fixed all the parameters, and only allowed the maximum disease transmission rate $\lambda_0$ to vary.

Figure~\ref{fig2} demonstrates the dependence of the transmission rate $\lambda$ and the natural plant growth rate $r$ on the rate of awareness $\sigma$, when computed at the endemic steady state $E^*$. One observes that as farmers become more aware of the spreading disease due to a higher level of $\sigma$, this leads to the larger steady state value of total awareness $M$ which, in turn, results in a larger level of application of nutrients and insecticides, which leads to the decrease in the steady state level of disease transmission $\lambda$ and a higher level of natural growth rate $r$. It should be noted, however, that this effect is only significant for lower values of $\sigma$, and further increase in awareness stemming from observing infected plants does not result in any significant changes in the disease transmission or the plant growth rate. To obtain a better insight into the effects of awareness on the endemic steady state, in Fig.~\ref{fig3} we plot the steady state values of the healthy and infected plant populations depending on two awareness rates. As expected, higher values of $\omega_0$ and $\sigma$ correspond to higher values of the steady state healthy population, and a lower value of the infected population. For the particular choice of parameters in this Figure, the endemic steady state is stable for any combination of $\omega_0$ and $\sigma$ values shown.

\begin{figure}
\hspace{-0.8cm}
 \centering
  \includegraphics[width=6.5in]{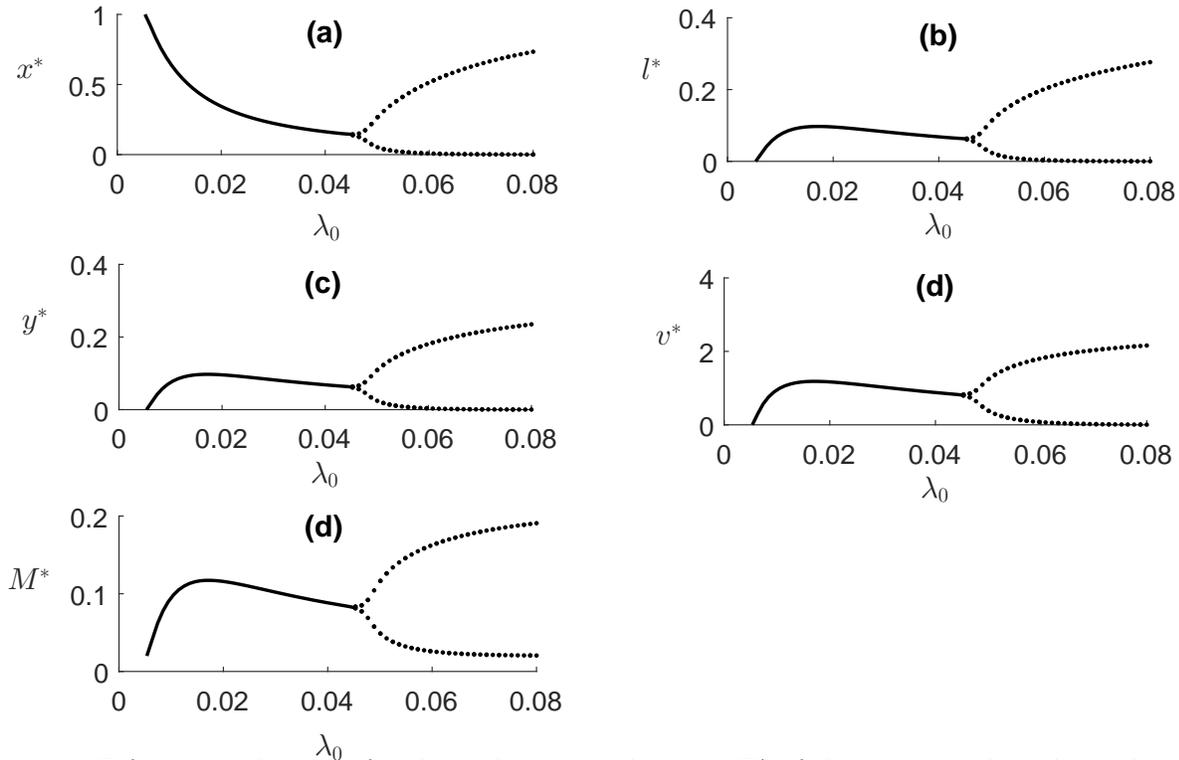}
  \vspace{-0.5cm}
  \caption{Bifurcation diagram for the endemic steady state $E^*$ of the system taking $\lambda_0$ with parameter values the same as in Fig.~\ref{fig1}. The steady state values of all populations are plotted and the minimum/maximum of the periodic solution when it exists.}\label{fig4}
  \vspace{-0.3cm}
\end{figure}

\begin{figure}
\hspace{-0.8cm}
\centering
\includegraphics[width=6in]{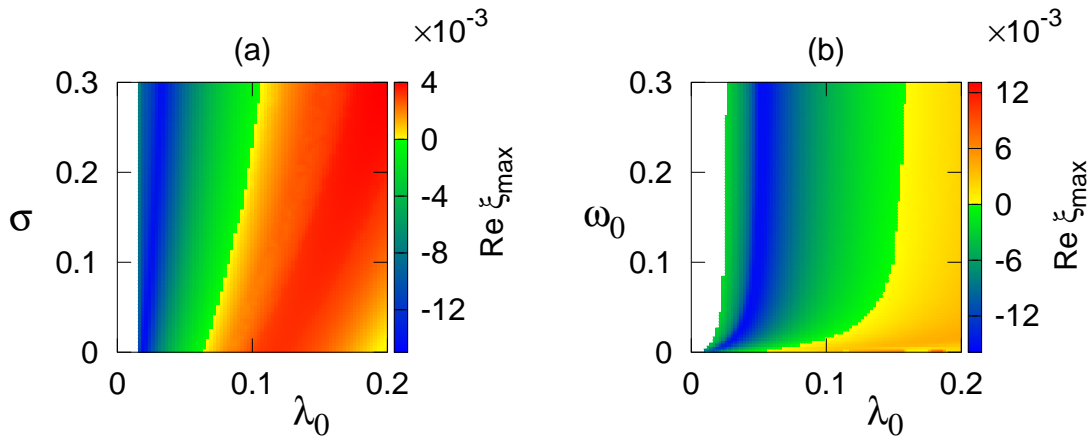}
\vspace{-0.2cm}
\caption{Stability of the endemic steady state $E^*$. Colour code denotes max$[Re(\xi)]$ whenever the endemic steady state is feasible. Parameter values are as follows, (a) $\omega_0=0.003$, (b) $\sigma=0.015$, and other parameters as in Figure~\ref{fig1}, except for $\tau=0.016$.}\label{fig5}
\vspace{-0.3cm}
\end{figure}

As one of the most important parameters characterising disease dynamics is the maximum disease transmission rate $\lambda_0$, Figure~\ref{fig4} shows a bifurcation diagram for the endemic steady state $E^*$ depending on $\lambda_0$. For very small values of $\lambda_0$, we have $R_0<1$, and, in agreement with {\bf Theorem~\ref{Thm1}}, the disease-fee steady state $E_1$ is stable, and the endemic steady state $E^*$ is not biologically feasible. As the value of $\lambda_0$ increases, the disease-free steady state loses its stability via a transcritical bifurcation, and a stable endemic steady state appears. For even larger values of $\lambda_0$, $E^*$ becomes unstable via Hopf bifurcation, giving rise to a stable periodic solution. Figure~\ref{fig4} also shows minima and maxima of this solution, suggesting that the amplitude of oscillations itself increases with $\lambda_0$, with the minimum values of populations on a periodic orbit being very close to zero. Figure~\ref{fig5} illustrates how stability of the endemic steady state depends on the relation between $\lambda_0$ and the two rates of awareness. One observes that for very small $\lambda_0$, the disease-free steady state is stable, and the endemic state is not feasible. Increasing $\lambda_0$ results in the emergence of a stable endemic steady state, while interestingly, further increase in $\lambda_0$  leads to a destabilisation of $E^*$ and the appearance of oscillations. As noted earlier in {\bf Remark~\ref{Rem1}}, for sufficiently small values of the disease transmission rate $\lambda_0$, it is possible to achieve disease eradication through a higher rate of global awareness campaigns $\omega_0$, while awareness arising from the observation of infected plants does not have such an effect.

\begin{figure}
\includegraphics[width=6.5in]{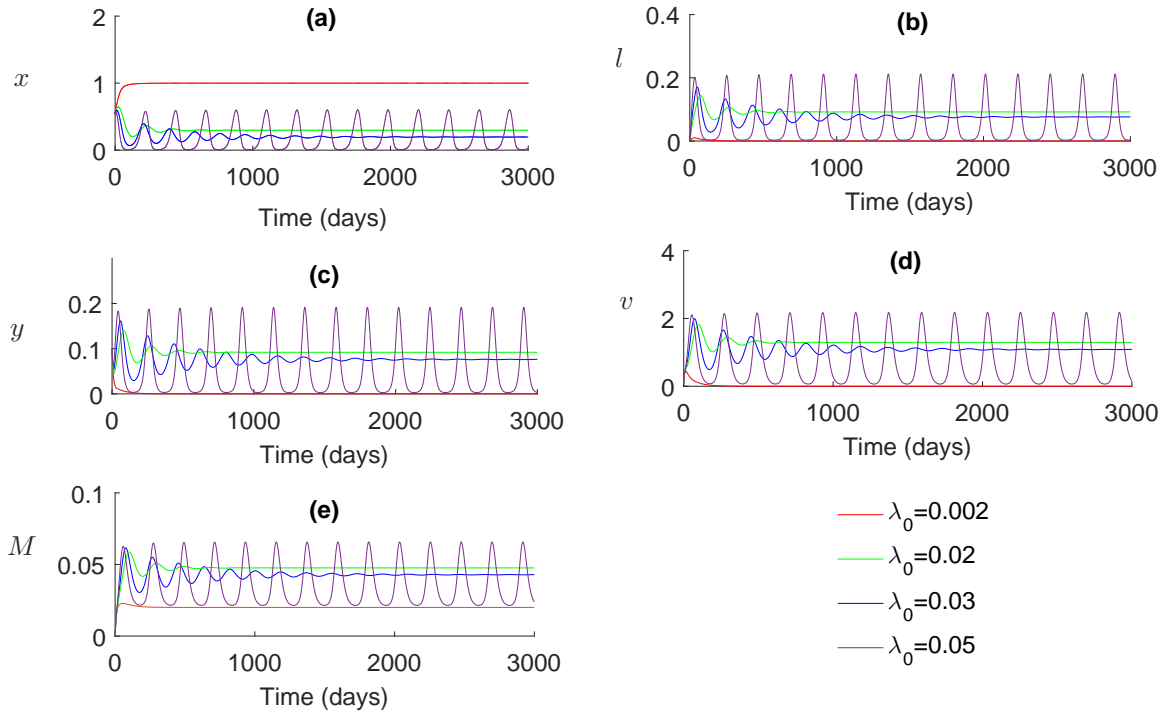}
\caption{Numerical solution of the system (\ref{model2}) with parameter values as in Fig.~\ref{fig1}: eradication of disease as signified by a stable disease-free steady state ($\lambda_0=0.002$, red); stable endemic steady state ($\lambda_0=0.02$, green, and $\lambda_0=0.03$, blue), and periodic oscillations around the endemic steady state ($\lambda_0=0.05$, purple).}\label{fig6}
\vspace{-0.3cm}
\end{figure}

Figure~\ref{fig6} illustrates different dynamical regimes that can be exhibited by the model, starting with a stable disease-fee steady state for small value of $\lambda_0$. For larger disease transmission rates, we observe the transition to a stable endemic steady state, with oscillatory approach to this steady state, suggesting that the largest characteristic eigenvalues are actually a pair of complex conjugate eigenvalues with a negative real part, which is increasing with $\lambda_0$. As $\lambda_0$ crosses a threshold for a Hopf bifurcation described in {\bf Theorem~\ref{Hopf_thm}}, the system settles on a stable periodic solution.

\section{Discussion}

In this paper we have analysed the dynamics of mosaic disease in the presence of human intervention in the form of applying nutrients and using insecticides depending on the level of population awareness about the disease. Analytical conditions for stability and bifurcations of the disease-fee and endemic equilibria have elucidated the role played by various parameters in determining the outcome of an epidemic, particularly from the perspective of awareness-based interventions. Interestingly, complete disease eradication, as characterised by a stable disease-free steady state, cannot be achieved purely by increasing rate of awareness arising from observation of infected plants. It can be done, though, by increasing the rate of awareness due to global media campaigns, which results in a higher value of population-level awareness at the disease-free steady state, and the associated higher level of using insecticides.

In terms of different dynamical regimes that can be exhibited by the model, our analysis and numerical calculations suggest that increasing the disease transmission rate results in an emergence of stable endemic steady state, which biologically corresponds to a sustained level of mosaic disease in the plant population. Further increase of the transmission rate leads to a destabilisation of this steady state and appearance of stable periodic oscillations. The intuitive explanation of these oscillations is as follows: higher transmission rate provides an opportunity for the growth of infected plant and vector populations; higher number of infected plants leads to an increased awareness, which, in turn, results in an increase in the use of nutrients and insecticides, which act to reduce the transmission rate and eliminate the infected vectors, and the cycle then repeats. An interesting and slightly counter-intuitive observation is that a higher rate disease transmission actually destabilises the endemic equilibrium, and there is an optimum intermediate range of values of transmission rate that allows the disease to maintain itself in the population without the risk of eradication or, for instance, stochastic extinction. It should be noted that although our model was developed for mosaic disease of Jatropha plant, the  results and conclusions are applicable to analysis of mosaic disease in other plants, such as cassava ({\it Manihot esculenta}), which is a major source of carbohydrates and a staple food in the developing world.

There are several directions in which the work presented in this paper could be extended. To simplify the analysis and make analytical progress, we modelled the process of vectors acquiring infection from infected plants indirectly, but it can be done more explicitly by separately considering the populations of uninfected and infected vectors, as in \cite{cmbn}. This would increase the dimensionality of the model, but potentially could provide better insights into the intricacies of the disease dynamics. Another interesting avenue to explore would be to allow not only continuous, but also pulse strategy for application of insecticide and use of nutrients, which better represents the way it is implemented in the field. Whilst we have analysed relative effects of different types of awareness and interventions, from a practical point of view it would also be important to study this as a control problem aimed at developing an optimal strategy for the use of nutrients and insecticides subject to various constraints, such as the minimum cost associated with campaigns and the use of chemicals, as well as minimum negative impact on environment, similar to \cite{eee}. The results of such analysis could then prove useful for design and implementation of policies for targeted awareness campaigns and control of mosaic disease.

\section*{Appendix A}\label{appendix}

The proof of Theorem~\ref{Hopf_thm} follows the methodology developed by Douskos and Markellos in \cite{Dous15}. The first step is to establish the conditions under which the characteristic polynomial has a pair of eigenvalues on the imaginary axis. This is provided by the following result, which is a slightly reformulated version of {\bf Lemma 2} and {\bf Theorem 3} from \cite{Dous15}.
\begin{lemma}\label{lem1}
The polynomial $H(\xi)=\xi^5+A_1\xi^4+A_2\xi^3+A_3\xi^2+A_4\xi+A_5$, $A_i\in\mathbb{R}$ $(i=1,\ldots,5)$, has one pair of purely imaginary roots $\xi_{1,2}= \pm i\sqrt{\theta}, \theta > 0$, and all other roots with non-zero real part if and only if the coefficients of this polynomial satisfy one of the following conditions
\[
\begin{array}{l}
(C_1):\quad (A_3-A_1A_2)(A_2A_5-A_3A_4)-(A_5-A_1A_4)^2=0,\\\\
\displaystyle{\theta=\theta_1=\frac{A_5-A_1A_4}{A_3-A_1A_2}>0,\quad A_3-A_1\theta_1\neq 0,}
\end{array}
\]
or
\[
\begin{array}{l}
(C_2):\quad A_5=A_1A_4, \quad A_3=A_1A_2, \quad A_4<0,\\\\
\displaystyle{\theta=\theta_2=\frac{1}{2}\left(A_2+\sqrt{A_2^2-4A_4}\right)>0,\quad A_1A_3\neq 0.}
\end{array}
\]
\end{lemma}
The proof of this lemma consists in identifying $\theta>0$, such that the original polynomial could be factorised as $H(\xi)=(\xi^2+\theta)g(\xi)$, thus proving the existence of a pair of purely imaginary roots $\xi_{1,2 }=\pm i\sqrt{\theta}$, while the conditions $A_3-A_1\theta\neq 0$ (respectively, $A_1A_3\neq 0$) ensure that all other roots for these parameter values have non-zero real parts \cite{Dous15}.

While {\bf Lemma~\ref{lem1}} establishes the existence of a pair of purely imaginary roots for some specific value of the chosen bifurcation parameter $\zeta=\zeta^*$, for a Hopf bifurcation to occur, there is an additional transversality condition
\[
\frac{d{\rm Re}(\xi)}{d\zeta}\Bigg|_{\zeta=\zeta^*}\neq 0,
\]
that needs to be satisfied to ensure that this pair of complex conjugate eigenvalues actually crosses the imaginary axis with a non-zero speed. According to an earlier result by Liu \cite{Liu}, this requirement is equivalent to verifying that $\Delta_4(\zeta^*)=0$ and $d\Delta_4(\zeta^*)/d\zeta\neq 0$, where $\Delta_4$ is the Hurwitz determinant associated with the Hurwitz matrix constructed from the coefficients of the characteristic equation: 
\[
\Delta_4(\zeta)=(A_3-A_1A_2)(A_2A_5-A_3A_4)-(A_5-A_1A_4)^2.
\]
According to {\bf Lemma~\ref{lem1}}, $\Delta_4(\zeta^*)=0$ in both cases $(C_1)$ and $(C_2)$. Orlando's formulas \cite{Gant} allow one to express this determinant in terms of roots $\xi_i$ of the characteristic polynomial $H(\xi)$ defined in {\bf Lemma~\ref{lem1}}
\[
\begin{array}{l}
\Delta_4(\zeta)=\Psi(\zeta)=(\xi_1+\xi_2)( \xi_1 +\xi_3)( \xi_1 +\xi_4)( \xi_1 +\xi_5)( \xi_2 +\xi_3)( \xi_2 +\xi_4)\\\\
\times ( \xi_2 +\xi_5)( \xi_3 +\xi_4)( \xi_3 +\xi_5)( \xi_4 +\xi_5).
\end{array}
\]
To show the connection between $d\Psi/d\zeta$ and $d{\rm Re}(\xi)/d\zeta$, let us assume that the condition $(C_1)$ in {\bf Lemma~\ref{lem1}} holds, and the polynomial $H(\xi)$ has a pair of complex conjugate eigenvalues $\xi_{1,2}$, such that
\[
\xi_{1,2}=\chi(\zeta)\pm i\nu(\zeta),
\]
with $\chi(\zeta^*)=0$, $\nu(\zeta^*)=\sqrt{\theta_1}$, and Re$(\xi_{3,4,5})\neq 0$. Substituting this into the definition of $\Psi(\zeta)$, differentiating
with respect to $\zeta$, and evaluating at $\zeta=\zeta^*$ yields
\[
\begin{array}{l}
\displaystyle{\frac{d\Psi(\zeta)}{d\zeta}\Bigg|_{\zeta=\zeta^*}=\Bigg[2(\nu^2+\xi_3^2)(\nu^2+\xi_4^2)(\nu^2+\xi_5^2)}\\\\
\hspace{2cm}\displaystyle{\times( \xi_3 +\xi_4)( \xi_3 +\xi_5)( \xi_4 +\xi_5)\frac{d\chi(\zeta)}{d\zeta}\Bigg]\Bigg|_{\zeta=\zeta^*}.}
\end{array}
\]
If all three remaining roots $\xi_3$, $\xi_4$, and $\xi_5$ have negative real parts at $\zeta=\zeta^*$, then
\[
\frac{d\chi(\zeta)}{d\zeta}\Big|_{\zeta=\zeta^*}\neq 0\Longleftrightarrow \frac{d\Psi(\zeta)}{d\zeta}\Big|_{\zeta=\zeta^*},
\]
On the other hand, assuming that not all of $\xi_3$, $\xi_4$, and $\xi_5$ have negative real parts at $\zeta=\zeta^*$, one can show \cite{Dous15} that it is not possible for any of the factors $(\xi_3 +\xi_4)$, $( \xi_3 +\xi_5)$, or $( \xi_4 +\xi_5)$ to be equal to zero, as this would violate the conditions of $(C_1)$. This completes the proof of the first part of {\bf Theorem~\ref{Hopf_thm}}. 

In the case where conditions $(C_2)$ hold, it immediately follows that $\Psi(\zeta^*)=0$, since $(A_3-A_1A_2)=(A_5-A_1A_4)=0$. In this case, to determine the sign of $d{\rm Re}(\xi)/d\zeta$ at $\zeta=\zeta^*$, we first note that the characteristic polynomial factorises as
\[
H(\xi,\zeta^*)=(\xi+A_1)(\xi^4+A_2\xi^2+A_4),
\]
which gives the roots as
\begin{equation}
\xi_{1,2}=\pm i\sqrt{\theta_2},\quad \xi_3= -A_1,\quad \xi_{4,5}=\theta_3.
\end{equation}
with $\theta_2=\frac{1}{2}\left(A_2+\sqrt{A_2^2+4|A_4|}\right)>0$, and $\theta_3=\frac{1}{2}\left(A_2-\sqrt{A_2^2+4|A_4|}\right)<0$. Vieta's formulas for the characteristic polynomial $H(\xi)$ allow one to relate coefficients of this polynomial to its roots $\xi_{1,2}=\chi(\zeta)\pm i\nu(\zeta)$, $\xi_3$, $\xi_4$ and $\xi_5$, as follows,
\[
\begin{array}{l}
A_1=-2\chi- (\xi_3 +\xi_4+\xi_5),\\
A_2=\chi^2+\nu^2+2\chi(\xi_3 +\xi_4+\xi_5)+\xi_3 \xi_4+\xi_3\xi_5+\xi_4\xi_5,\\
A_3=-(\chi^2+\nu^2)(\xi_3 +\xi_4+\xi_5)-2\chi(\xi_3 \xi_4+\xi_3\xi_5+\xi_4\xi_5)-\xi_3\xi_4\xi_5,\\
A_4=(\chi^2+\nu^2)(\xi_3 \xi_4+\xi_3\xi_5+\xi_4\xi_5)+2\chi\xi_3\xi_4\xi_5,\\
A_5=-(\chi^2+\nu^2)\xi_3\xi_4\xi_5.
\end{array}
\]

Differentiating these relations with respect to $\zeta$, evaluating them at $\zeta=\zeta^*$, and using the relations $\xi_4=-\xi_5$, $A_4=-\xi_4^2<0$ valid at $\zeta=\zeta^*$, gives
\[
\frac{d\chi(\zeta)}{d\zeta}\Bigg|_{\zeta=\zeta^*}=\left\{-\frac{A_1'\theta_2^2+(A_1A_2'-A_3')\theta_2-(A_1A_4'-A_5')}{2(\theta_2+A_1^2)(\theta_2+\xi_5^2)}\right\}\Bigg|_{\zeta=\zeta^*},
\]
which, with $\theta_2>0$, shows that
\[
\frac{d\chi(\zeta)}{d\zeta}\Bigg|_{\zeta=\zeta^*}\neq 0~~\Longleftrightarrow~~ \Big[A_1'\theta_2^2+(A_1A_2'-A_3')\theta_2-(A_1A_4'-A_5')\Big]\Big|_{\zeta=\zeta^*}\neq 0,
\]
thus completing the proof of {\bf Theorem~\ref{Hopf_thm}}.


\begin{thebibliography}{99}

\bibitem{ach} W.M.J. Achten, E. Mathijs, L. Verchot, V.P. Singh, R. Aerts, B. Muys, {\it Jatropha} biodiesel fueling sustainability? Biofuels, Bioprod.
Bioref. 1 (2007) 283-291.

\bibitem{Sah} N.K. Sahoo, A. Kumar, S. Sharma, S.N. Naik, Interaction of {\it Jatropha curcas} plantation with ecosystem, in Proc.
Int. Conf. Energy Env. (2009) 19-21.

\bibitem{Jong} R.E.E. Jongschaap, W.J. Corr\'e, P.S. Bindraban, W.A. Brandenburg, Claims and facts on {\it Jatropha curcas} L., Plant Research International, Report 158 (2007).

\bibitem{mang08} S. Mangkoedihardjo, Surahmaida, {\it Jatropha curcas} L. for phytoremediation of lead and cadmium polluted soil. World Appl. Sci. J. 4 (2008) 519-522.

\bibitem{pand12} V.C. Pandey, K. Singh, J.S. Singh, A. Kumar, B. Singh, R.P. Singh, {\it Jatropha curcas}: A potential biofuel plant for sustainable environmental development. Renew. Sustain. Energy Rev. 16 (2012) 2870-2883.

\bibitem{security} R.N. Strange, P.R. Scott, Plant disease: a threat to global food security, Ann. Rev. Phytopath. 43 (2005) 83-116.

\bibitem{Africa} O.J. Alabi, P. Lava Kumar, R.A. Naidu, Cassava mosaic disease: a curse to food security in sub-Saharan Africa, APSnet Feature, 2011.

\bibitem{virus diseases} J.M. Thresh, The impact of plant virus diseases in developing countries, in G. Loebenstein, G. Thottappilly (Eds.), Virus and Virus-like Diseases of Major Crops in Developing Countries, Springer, Amsterdam, 2003, pp. 1-30.

\bibitem{Kash} B.D. Kashina, M.D. Alegbejo, O.O. Banwo, S.L. Nielsen, M. Nicolaisen M. Molecular identification of a new begomovirus associated with mosaic disease of {\it Jatropha curcas} L. in Nigeria, Arch. Virol. 158 (2013) 511-514.

\bibitem{Nar} D.S. Aswatha Narayana, K.S. Shankarappa, M.R. Govindappa, H.A. Prameela, M.R. Gururaj Rao, K.T. Rangaswamy, Natural occurrence of Jatropha mosaic virus disease in India, Curr. Sci. 91 (2006) 584-586.

\bibitem{Gao} S.Q. Gao, J. Qu, N.-H. Chua, J. Ye, A new strain of {\it Indian cassava mosaic virus} causes a mosaic disease in the biodiesel crop {\it Jatropha curcas}, Arch. Virol. 155 (2010) 607-612.

\bibitem{Reddy} D.S. Aswatha Narayana, K.T. Rangaswamy, K.S. Shankarappa, M.N. Maruthi, C.N. Lakshminarayana Reddy, A.R. Rekha, K.V. Keshava Murthy, Distinct begomoviruses closely related to cassava mosaic virus cause indian Jatropha mosaic disease, Int. J. Virol. 3 (2007) 1-11.

\bibitem{Bedford} I.D Bedford, R.W. Briddon, J.K. Brown, R.C. Rosell, P.G. Markham, Geminivirus transmission and biological characterisation of {\it Bemisia tabaci} (Gennadius) biotypes from different geographic regions, Ann. Appl. Biol. 125 (1994) 311-325.

\bibitem{Fauq} C. Fauquet, D. Fargette, African cassava mosaic virus: etiology, epidemiology and control, Plant Dis. 74 (1990) 404-411.

\bibitem{Fargette} D. Fargette, M. Jeger, C. Fauquet, L.D.C. Fishpool, Analysis of temporal disease progress of African cassava mosaic virus, Phytopath. 84 (1994) 91-98.

\bibitem{strategy} J.M. Thresh, G.W. Otim-Nape, Strategies for controlling African cassava mosaic Geminivirus, in K.F. Harris (Ed.), Advances in Disease Vector Research, Springer, New York, 1994, pp. 215-236.

\bibitem{strategy2} S.E. Seal, M.J. Jeger, F. Van den Bosch, Begomovirus evolution and disease management, Adv. Virus Res. 67 (2006) 297-316.

\bibitem{Ahohuendo} B.C. Ahohuendo, S. Sarkar, Partial control of the spread of African cassava mosaic virus in Benin by intercropping, J. Plant Dis. Prot. 102 (1995) 249-256.

\bibitem{soap} G.D. Butler, Jr., T.J. Henneberry, P.A. Stansly, D.J. Schuster, Insecticidal effects of selected soaps, oils and detergents on the sweetpotato whitefly: (Homoptera: Aleyrodidae), Florida Entomol. 76 (1993) 161-167.

\bibitem{Thresh} J.M. Thresh, R.J. Cooter, Strategies for controlling cassava mosaic virus disease in Africa, Plant Path. 54 (2005) 587-614.

\bibitem{Butler} G.D. Butler, Jr., T.J. Henneberry, Cottonseed oil and safer insecticidal soap: effects on cotton and vegetable pests and phytotoxicity,
Southwest. Entomol. 15 (1990) 257-264.

\bibitem{Opa} A.M. Oparaeke, M.C. Dike, C.I. Amatobi, Insecticidal efficacy of SABRUKA formulations as protectants of cowpea against field pests, J. Entomol. 3 (2006) 130-135.

\bibitem{nutrient} C. Dordas, Role of nutrients in controlling plant diseases in sustainable agriculture: a review, in E. Lichtfouse, M. Navarrete, P. Debaeke, S. V\'eronique, C. Alberola (Eds.), Sustainable agriculture, Springer, Amsterdam, 2009, pp. 443-460.

\bibitem{nutrient1} S. Pennazio, P. Roggero, Mineral nutrition and systemic virus infections in plants, Phytopath. Mediter. 36 (1997) 54-66.

\bibitem{1970} R. Singh, Influence of nitrogen supply on host susceptibility to Tobacco mosaic virus infection, Phyton 14 (1970) 37-39.

\bibitem{Graham1} R.D. Graham, Effects of nutrient stress on susceptibility of plants to disease with particular reference to the trace elements, Adv. Bot. Res. 10 (1983) 221-276.

\bibitem{Graham2} R.D. Graham, M.J. Webb, Micronutrients and disease resistance and tolerance in plants, in: J.J. Mortvedt, F.R. Cox, L.M. Shuman, R.M. Welch (Eds.), Micronutrients in Agriculture, 2nd ed., Soil Science Society of America, Madison, Wisconsin, USA, 1991, pp. 329-370.

\bibitem{Huber1999} D.M. Huber, R.D. Graham, The role of nutrition in crop resistance and tolerance to diseases, in: Z. Rengel (Ed.), Mineral nutrition of crops: fundamental mechanisms and implications, Food Product Press, New York, 1999, pp. 169-204.

\bibitem{Schut} M. Schut, J. Rodenburg, L. Klerkx, A. van Ast,  L. Bastiaans, Systems approaches to innovation in crop protection. A systematic literature review, Crop Prot. 56 (2014) 98-108.

\bibitem{Klerkx} L. Klerkx, N. Aarts, C. Leeuwis, Adaptive management in agricultural innovation systems: The interactions between innovation networks and their environment, Agricult. Syst. 103 (2010) 390-400.

\bibitem{aware_knowledge1} W. El Khoury, K. Makkouk, Integrated plant disease management in developing countries, J. Plant Path. 92 (2010) S35-42.

\bibitem{aware_plant1} G.L. Schumann, C.J. D'Arcy, Plant pathology courses for agricultural awareness, Plant Dis. 83 (1999), 492-501.

\bibitem{Khan} G.A. Khan, S. Muhammad, K. Mahmood Ch., M.A. Khan, Information regarding agronomic practices and plant protection measures obtained by the farmers through electronic media, J. Anim. Plant Sci. 23 (2013) 647-650.

\bibitem{LeB} F. Le Bellec, A. Rajaud, H. Ozier-Lafontaine, C. Bockstaller, E. Malezieux, Evidence for farmers' active involvement in co-designing citrus cropping systems using an improved participatory method, Agron. Sust. Devel. 32 (2012) 703-714.

\bibitem{Kumar2} M. Kumar, I.J. Kuppast, K.L. Mankani, K. Chandra Prakash, T. Veershekar, Shekhashavali. Use and awareness of pesticides in Malnad region of Karnataka, J. Pharm. Res. 5 (2012) 3875-3877.

\bibitem{Yang2014} X. Yang, F. Wang, L. Meng, W. Zhang, L. Fan, V. Geissen, C.J. Ritsema, Farmer and retailer knowledge and awareness of the risks from pesticide use: A case study in the Wei River catchment, China, Sci. Total Env. 497-498 (2014) 172-179.

\bibitem{school} G. Feder, R. Murgai, J.B. Quizon, Sending farmers back to school: The impact of farmer field schools in Indonesia, Appl. Econ. Persp. Policy 26 (2004) 45-62.

\bibitem{school2} G. Feder, R. Murgai, J.B. Quizon, The acquisition and diffusion of knowledge: The case of pest management training in farmer field schools, Indonesia. J. Agricult. Econ. 55 (2004) 221-243.

\bibitem{field_schools} A.R. Braun, G. Thiele, M. Fern\'andez, Farmer field schools and local agricultural research committees: complementary platforms for integrated decision-making in sustainable agriculture. Overseas Development Institute, AGREN Network Paper No. 105, 2000.

\bibitem{Mogaka} V. Mogaka, A. Ehrensperger, M. Iiyama, M. Birtel, E. Heim, S. Gmuender, Understanding the underlying mechanisms of recent {\it Jatropha curcas} L. adoption by smallholders in Kenya: A rural livelihood assessment in Bondo, Kibwezi, and Kwale districts, Energy Sust. Dev. 18 (2004), 9-15.

\bibitem{Favr} N. Favretto, L.C. Stringer, A.J. Dougill, Towards improved policy and institutional coherence in the promotion of sustainable biofuels in Mali. Env. Policy and Gov. 25 (2015) 36Ð54.

\bibitem{WRM08} World Rainforest Bulletin, Issue 137 (2008).

\bibitem{aware_mosaic1} E. Moses, Development of appropriate strategies to control cassava diseases in Ghana, in The Role of Plant Pathology in Food Safety and Food Security, 2009, Springer, Berlin, pp. 11-24.

\bibitem{aware_mosaic2} M.M. Chipeta, P. Shanahan, R. Melis, J. Sibiya, I.R. Benesi, Farmers' knowledge of cassava brown streak disease and its management in Malawi, Int. J. Pest Manag. 62 (2016) 175-184.

\bibitem{Misra1}  A.K. Misra, A. Sharma, J.B. Shukla, Modeling and analysis of effects of awareness programs by media on the spread of infectious diseases, Math. Comput. Model. 53 (2011) 1221-1228.

\bibitem{Misra2} A.K. Misra, P.K. Tiwari, E. Venturino, Modeling the impact of awareness on the mitigation of algal bloom in a lake, J. Biol. Phys.  42 (2016) 147-165.

\bibitem{ManOn} P. Manfredi, A. d'Onofrio, Modeling the interplay between human behavior and the spread of infectious diseases, Springer, New York, 2013.

\bibitem{Cui} J. Cui, Y. Sun, H. Zhu, The impact of media on the control of infectious diseases, J. Dyn. Diff. Eqns. 20 (2008) 31-53.

\bibitem{Funk} S. Funk, M. Salath\'e, V.A.A. Jansen, Modelling the influence of human behaviour on the spread of infectious diseases: a review, J. R. Soc. Interface 7 (2010) 1247-1256.

\bibitem{agaba1} G.O. Agaba, Y.N. Kyrychko, K.B. Blyuss, Mathematical model for the impact of awareness on the dynamics of infectious diseases, Math. Biosci. 286 (2017) 22-30.

\bibitem{greenh} D. Greenhalgh, S. Rana, S. Samanta, T. Sardar, S. Bhattacharya, J. Chattopadhyay, Awareness programs control infectious disease - multiple delay induced mathematical model, Appl. Math. Comp. 251 (2015) 539-563.

\bibitem{zuo} L. Zuo, M. Liu, J. Wang, The impact of awareness programs with recruitment and delay on the spread of an epidemic, Math. Prob. Eng. 2015 (2015) 235935.

\bibitem{agaba2} G.O. Agaba, Y.N. Kyrychko, K.B. Blyuss, Time-delayed SIS epidemic model with population awareness, Ecol. Compl. 31 (2017) 50-56.

\bibitem{agaba3} G.O. Agaba, Y.N. Kyrychko, K.B. Blyuss, Dynamics of vaccination in a time-delayed epidemic model with awareness, Math. Biosci. 294 (2017) 92-99.

\bibitem{mmas} F. Al Basir, E. Venturino, P.K. Roy, Effects of awareness program for controlling mosaic disease in Jatropha curcas plantations, Math. Meth. Appl. Sci. 40 (2017) 2441-2453.

\bibitem{ABR} F. Al Basir, P.K. Roy, Dynamics of mosaic disease with roguing and delay in Jatropha curcas plantations, J. Appl. Math. Comput. (2017) in press.

\bibitem{cmbn} P.K. Roy, X.-Z. Li, F. Al Basir, A. Datta, J. Chowdhury, Effect of insecticide spraying on {\it Jatropha curcas} plant to control mosaic virus: a mathematical study, Commun. Math. Biol. Neurosci. 2015 (2015) 36.

\bibitem{eee} E. Venturino, P.K. Roy, F. Al Basir, A. Datta,  A model for the control of the mosaic virus disease in {\it Jatropha curcas} plantations, Energy, Ecol. Env. 1 (2016) 360-369.

\bibitem{Vir} H. Czosnek, A. Hariton-Shalev, I. Sobol, R. Gorovits, M. Ghanim, The  incredible journey of begomoviruses in their whitefly vector, Viruses 9 (2017) 273.

\bibitem{Holt97} J. Holt, M.J. Jeger, J.M. Thresh, G.W. Otim-Nape, An epidemilogical model incorporating vector population dynamics applied to African cassava mosaic virus disease, J. Appl. Ecol. 34 (1997) 793-806.

\bibitem{Jeger04} M.J. Jeger, J. Holt, F. van den Bosch, L.V. Madden, Epidemiology of insect-transmitted plant viruses: modelling disease dynamics and control interventions, Physiol. Entomol. 29 (2004) 291-304.

\bibitem{Jackson} M. Jackson, B.M. Chen-Charpentier, Modeling plant virus propagation with delays, J. Comp. Appl. Math. 309 (2017) 611-621.

\bibitem{Misra11} A.K. Misra, A. Sharma, and V. Singh, Effect of awareness programs in controlling the prevalence of an epidemic with time delay, J. Biol. Syst. 19 (2011) 389-402.

\bibitem{nutri} K.J. Boote, J.W. Jones, N.B. Pickering, Potential uses and limitations of crop models, Agron. J. 88 (1996) 704 - 716.

\bibitem{Dous15} C. Douskos, P. Markellos, Complete coefficient criteria for five-dimensional Hopf bifurcations, with an application to economic dynamics, J. Nonlin. Dyn. 2015 (2015) 278234.

\bibitem{Liu} W.-M. Liu, Criterion of Hopf bifurcations without using eigenvalues, J. Math. Anal. Appl. 182 (1994) 250-256.

\bibitem{Gant} F.R. Gantmacher, Applications of the theory of matrices, Interscience Publishers, New York (1959).

\end{thebibliography}
\end{document}